# Title:

# A complementary method for automated detection of microaneurysms in fluorescein angiography fundus images to assess diabetic retinopathy

Authors:

1-      Meysam Tavakoli
Ph.D. Student in Physics Department, Oklahoma State University; Research assistant, Center for research in Medical Physics and Biomedical Imaging from Mashhad University of Medical Sciences (MUMS) and member of Eye Image Analysis Research Group between MUMS and Ferdowsi University of Mashhad.
Tel/fax: +1 405 744-5812
2-      Reza Pourreza Shahri
Ph.D. Candidate, Electrical Engineering Department, University of Texas at Dallas. Research assistant of Eye Image Analysis Research Group between MUMS and Ferdowsi University of Mashhad.
Tel/fax: +989153257806
3-      Hamidreza Pourreza
Associate Professor, Computer Engineering Department, Ferdowsi University of Mashhad, Mashhad, Iran. Research assistant of Eye Image Analysis Research Group between MUMS and Ferdowsi University of Mashhad.
Tel: +985118815100 (407)
4-      Alireza Mehdizadeh
Assistant professor of Medical Physics, Center for research in Medical Physics and biomedical Engineering-Image processing lab., Shiraz University of Medical Sciences, Shiraz- Iran.
Tel/fax: +989371385883
5-      Touka Banaee
Associate Professor, Ophthalmic Research Center, Khatam-Al-Anbia Hospital, Mashhad University of Medical Sciences, Research assistant of Eye Image Analysis Research Group between MUMS and Ferdowsi University of Mashhad.
Tel/fax: +989153100045
6-      Mohammad Hosein Bahreini Toosi
Professor of Medical Physics, Mashhad University of Medical Sciences, Department of Medical Physics, School of Medicine.
Tel/fax: +985118828576

**Corresponding author:**
Meysam Tavakoli

Dept. of Physics, PS-145- Oklahoma State University
Stillwater, Oklahoma, USA
Tel: +1 405 744-5812

Email: MeysamTavakkoli@gmail.com

        Meysam.Tavakoli@okstate.edu

**Abstract**—Early detection of microaneurysms (MAs), the first sign of Diabetic Retinopathy (DR), is an essential first step in automated detection of DR to prevent vision loss and blindness. This study presents a novel and different algorithm for automatic detection of MAs in fluorescein angiography (FA) fundus images, based on Radon transform (RT) and multi-overlapping windows. This project addresses a novel method, in detection of retinal land marks and lesions to diagnose the DR. At the first step, optic nerve head (ONH) was detected and masked. In preprocessing stage, top-hat transformation and averaging filter were applied to remove the background. In main processing section, firstly, we divided the whole preprocessed image into sub-images and then segmented and masked the vascular tree by applying RT in each sub-image. After detecting and masking retinal vessels and ONH, MAs were detected and numbered by using RT and appropriated thresholding. The results of the proposed method were evaluated reported on three different retinal images databases, the Mashhad Database with 120 FA fundus images, Second Local Database from Tehran with 50 FA retinal images and a part of Retinopathy Online Challenge (ROC) database with 22 images. Automated DR detection demonstrated a sensitivity and specificity of 94% and 75% for Mashhad database and 100% and 70% for the Second Local Database respectively.

## 1. Introduction

Diabetic retinopathy (DR) is the single largest cause of sight loss and blindness in the working age population of western countries [1]. Moreover, world health organization (WHO) anticipates that the total number of diabetic patients will at least reach to 366 million by 2030 [2]. According to some estimations, more than 75% of patients with diabetes within 15 to 20 years of diagnosis diabetes are threatened by DR [3]. These are likely affected by diabetes in their retina that is an important threat to vision.

Generally, the probability of blindness in a person with diabetes is 25 times more than the general population, but by early detection of DR and timely treatment, visual loss and blinding can be prevented [4]. Since early diagnosis of DR is critical in preventing vision loss, annual ophthalmic examinations are suggested by the American Academy of Ophthalmology and the American Diabetes Association. Due to high burden of diabetic patients, high cost of examination, and shortage of ophthalmologists, especially in rural areas, automated systems can help in saving the time of ophthalmologists and enhancing the tele-medical diagnosis of the condition.

Automated Computer Diagnosis (ACD) and Computer Aided Diagnosis (CAD) systems have a potential to analyze large number of images with saving time and money. Such automated systems are able to identify early signs of DR and diagnose based on some criteria defined by the ophthalmologists; moreover, could reduce the workload of them by 50% [5]. In addition, these systems would allow ophthalmologists to check more patients and spend more time for whom; they are actually in need of their expertise. On the other hand, it has been found that one of the most important sources of expenditure in setting up any DR detection program is related to the cost of learning trained manual readers. As a result, finding the potential ability of Automated Detection of Diabetic Retinopathy (ADDR) can play a significant role in diminishing this cost. If automated detection programs are capable to reject a large number of those patients who do not have DR, it will reduce the workload of the trained readers and thus reduce the costs. Other benefits of an automated detection program include release from fatigue and improved repeatability.

The first signs of DR are Microaneurysms (MAs), tiny dilations of the capillaries that do not affect vision. Moreover, MAs detection is the most important step in the automated detection of DR systems. MAs are visible immediately after the arterial phase of fluorescein angiography (FA), a technique used to evaluate the circulation in the retina and choroid. MAs counting has been used as a tool for determination of the progression of the DR and there are many protocols that perform this task [6]. A number of different methods for automated MA detection have been proposed in the past (see section 2). Nevertheless, these methods were tested on different databases using various evaluation measures that made it impossible to direct comparison of performance [7] .

## 2. State of the Art

Since the detection of MA is so important for ACD and CAD of DR, there are a large number of studies about this subject.

Lots of publications focused on color retinal images for MA detection [7-16]. Most of significant lesions due to DR are visible in color fundus images; however, they cannot be utilized as gold standard for DR diagnosis. The reason is that MAs often are not readily visualized and may be confused with small dot hemorrhages in the color images (accurate

MAs detection and counting used as an important tool for determination of the DR progression). Using FA technique increases the visibility of these lesions. Moreover, some MAs may emerge in connection with larger vessels, or may be a part of an accumulation of more than one MA. In addition, small MAs can look similar to other retinal pathologies or capillary crossings. Therefore, the localization of the MAs to count is not easy. For detection of MAs, FA images allow with much higher sensitivity and considered as gold standard procedure. MAs are visible as bright spots with nice contrast in FA, and it is possible to distinguish them from other features or lesions such as small vessels by using morphological processing and structuring elements in the thresholded image [17].

The earliest published papers in MAs detection belong to Baudoin *et al*. [18] and Lay [19]. Baudoin *et al* described a mathematical morphology based detection approach for MAs in fluorescein angiograms. Lay presented an algorithm to detect MA in FA in 1983 [19]. In order to remove the MA and preserve the linear vessels, he applied morphological openings with linear structuring elements in different directions. The subtracted image (top-hat transformation) showed details which may correlate to MAs.

Spencer *et al*. [20] utilized shade correction and normalization methods for increasing dynamic range. After that MA candidates were specified by means of a matched filter technique. Moreover, they applied a morphological operation to detect MAs in FA images in another study [21].

Mendonça *et al*. [22] worked on FA retinal image for detection of MAs. Their algorithm included preprocessing and enhancement steps as well as segmentation of MAs.

Walter *et al*. [23] worked on FA fundal images for detection of MAs by using top-hat transformation, matching filter and automatic thresholding. Furthermore, they used a registration algorithm for monitoring images coming from different examinations.

Cree *et al*. [24] reestablished the region growing and classification algorithm for selecting region of interest, developed by Spencer *et al*. [21]. In their method, MAs were counted and an automated procedure for registration was done to allow sequential comparisons of MAs.

Hipwell *et al*. [25] used the same method of Spencer *et al*. [20] followed on $50^{°}$ digital red-free images, by removing vessels and other suspectors by the top-hat transformation and use of the Gaussian filter to reserve candidate MAs.

Kamel *et al*. [26] applied neural network structure for detection of MAs in retinal angiograms. The learning vector quantization neural network was used to classify the input patterns into their fancy classes using competitive layers.

Hafez and Azeem, applied Canny edge detector to find and remove vessel segments after that Sobel edge detection was used for the neighborhoods of the remaining objects [27].

Goatman *et al*. [28] provided an ADDR program for detection MAs in FA images with the ability to follow temporal changes in MA turnover. They used the same technique as Cree *et al*. [24] for MAs detection and image registration.

Most of presented studies applied directly the algorithm based on an image processing technique called Mathematical Morphology. On the other hand, related to our study there is one study that used Radon cliff to detect MAs in color images [29].

The main purpose of our work is to develop "automatic complementary system" for detection of early vascular DR lesions, MAs, in retinal FA images. Our algorithm introduces combination of morphological and non-morphological methods in detection of MAs in fluorescein angiographic fundus images.

### 3. Proposed Method

A total number of 170 FA fundus images from two local sources were captured (140 with DR and 30 without DR) from both left and right eyes of people. Images in each dataset were divided into training and test sets.

The first set was provided by Khatam-Al-Anbia Eye Hospital, named as MUMS-DB (**M**ashhad **U**niversity of **M**edical **S**ciences **D**atabase). This set consisted of 100 cases with DR, in different stage of DR as case group, and 20 without DR, with no systemic disease or ocular micro-vascular involvement as control group. Image distribution does not correspond to any typical population, i.e. the data is biased and no priori information can be derived from it. All images were taken with a 50° field of view a Canon TRC-50EX Mydriatic retinal camera under the same lighting conditions by same photographer. Image resolution was $2896 \times 1944$ pixels Tag Image File Format (TIFF).

In addition, to evaluate the efficiency of our algorithm, we tested it with the Second Local Database ($2^{nd}$-DB), from Tehran. The $2^{nd}$-DB consisted of 50 FA retinal images. This database was taken with a 45° field of view of TOPCON (TRC-NW100) Mydriatic retinal camera with image resolution of $640 \times 480$ pixels TIFF. This set consisted of 40 cases with DR and 10 non-DR fundus images.

Ground truth file used in this project was in image format prepared by a vitreoretinal specialist and a trained ophthalmologist who marked MAs separately on a transparent layer with similar dimension for each retinal image in train and test set. Therefore, this file not only contains the number of MAs, but also shows exact location of each MA in every image, which is used in lesion based analysis. It should be mentioned that there were some images with blurriness or coagulations in our image dataset and we did not omit them.

Moreover, we tested our approach on 22 images selected from the Retinopathy Online Challenge (ROC) database [7]. These images were converted to gray levels and used in our study, with resolution of 768×576 at the default JPEG compression settings.

The overall schematic of our method is shown in Fig. 1.

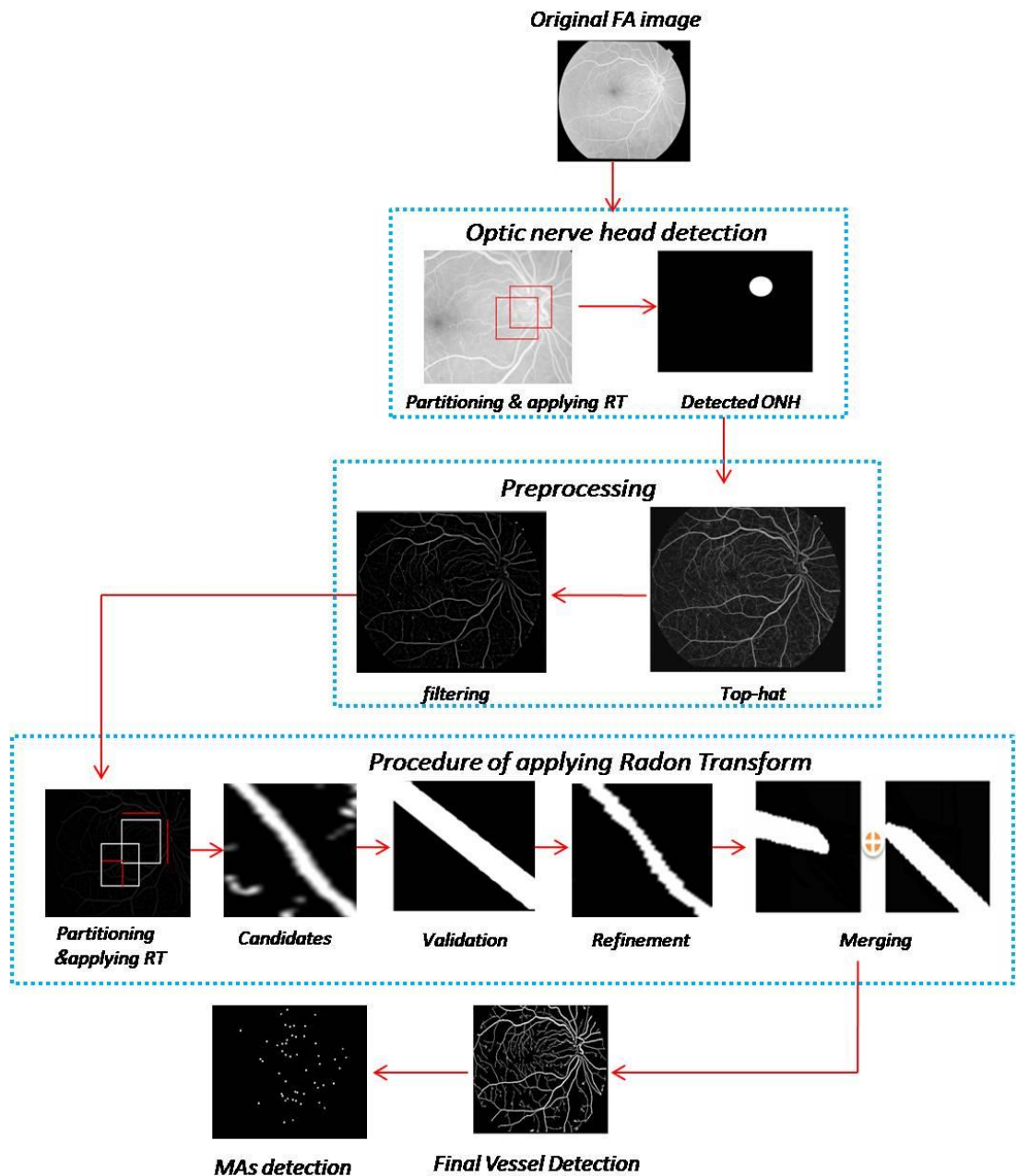

**Fig. 1.** Block diagram of proposed method

*3.1.    Multi-overlapping window*

In the proposed algorithm fundus image was partitioned into widows or sub-images. To determine the size of each sub-image, we used our knowledge database. In this regard, size of targeted objects specified the size of the windows (*n*). Besides, to find objects on border of sub-images, we have to define an overlapping factor of sliding windows (*step*). If window's *step* is equal to one, we will search every pixel of images just one time and sub-images only touch each other without any overlapping and if *step* is defined as two or more, then we go (*n/step*) pixel each time either in horizontal or vertical sliding then each pixel would belong to up to $n^2$ sub-images. In other words, the sliding window has an overlapping factor which means that the neighbor windows are not separated and are overlapped. Window moves in any direction in the image horizontally and vertically. The Fig. 2 shows the procedure of sliding window in the original image, the (*n*) and (*step*).

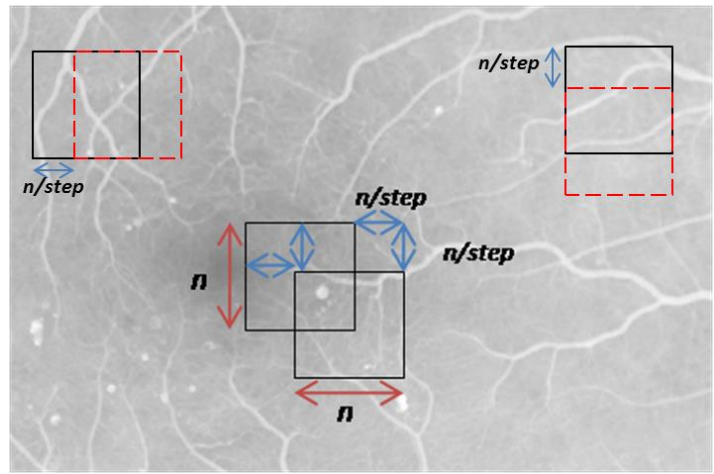

**Fig. 2.** Window size and overlapping ratio (n, step).

Beside applying multi-overlapping window in this study, Radon transform was used for MAs detection.

### 3.2.   *Radon Transform (RT)*

RT is at the center of mathematical model for measurements made in X-ray computed tomography (CT). This model is later inversed and used to reconstruct images of organs inside the human body.

RT of a continuous two dimensional function *f* is defined by:

$$\breve{f}_\theta(s) = \int_{-\infty}^{+\infty} \int_{-\infty}^{+\infty} f(x,y)\delta(s - x\cos\theta - y\sin\theta) \ dx \, dy \qquad (1)$$

In equation (1), $\delta$ denotes the Dirac delta function which picks out the path of the line whose distance to the origin is *s* and the angle of its normal vector with the *x* axis is *θ*. A simple geometric interpretation of RT is shown in Fig. 3.

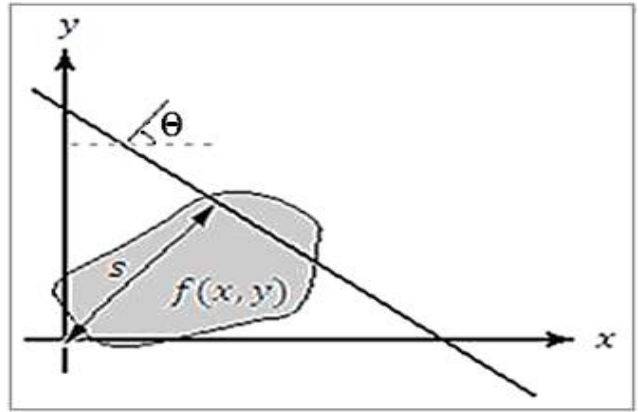

**Fig. 3.** Radon transform.

Equation (1) can be expressed in a single integral as:

$$\breve{f}_\theta(s) = \int_{-\infty}^{+\infty} f(x,y)ds = \int_{-\infty}^{+\infty} f\big(s(\cos\theta + \sin\theta) + z(-\sin\theta + \cos\theta)\big)ds \qquad (2)$$

Where

$$\begin{bmatrix} x \\ y \end{bmatrix} = \begin{bmatrix} \cos\theta & \sin\theta \\ \sin\theta & \cos\theta \end{bmatrix} \begin{bmatrix} s \\ z \end{bmatrix}$$

In which the z-axis, in Fig. 4, is along the line in Fig.3.

The quantity $\breve{f}_\theta(s)$ in equation (1) may be interpreted as the one-dimensional function of a single variable *s* with *θ* as a parameter, and with the arrangement of Fig. 4 this $\breve{f}_\theta(s)$ is referred to as a parallel projection.

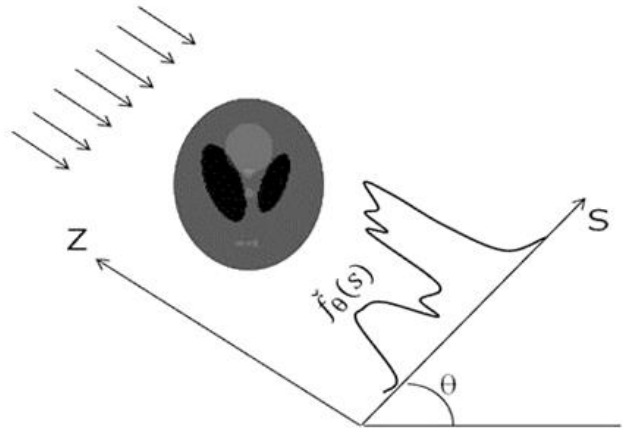

**Fig. 4.** Parallel projection line integral

RT is integral of an image over straight lines in specified angle. It makes our algorithm sturdy and less sensitive to noise than other algorithms because intensity variations due to noise tend to be omitted by process of integration.

In spite of morphologic operators which are image based and could not provide numerical data directly, by using RT object finding and description will be done simultaneously.

RT is able to transform line-containing images into a domain where each line in the image gives a peak or a valley in Radon domain. For line detection some numerical data such as line width, and line angle could be achieved directly from Radon matrix.

### 3.3. Detection of Optic Nerve Head (ONH)

The ONH is characterized as a bright circular object set against a background and because of large vessels coming out and going in through it; its edges are ill defined. The ONHs are non-uniform in intensity, size, and location.

Our algorithm, for ONH detection, was composed of 4 steps:

1. Generation of sub-images

2. Applying  Radon transform

3. ONH certifying

4. ONH mask

At first, the ONH should be extracted in local windows. The window size ($n$) has a direct effect on the extraction accuracy. Based on our knowledge about the ONH size, appropriate size of window, $n$, was chosen equal to maximum diameter of the biggest ONH in pixel in our databases ($n$=309 pixels for MUMS-DB, $n$=38 pixels for $2^{nd}$-DB and $n$=114 pixels for ROC database). The windows overlapping ratio was another important parameter which affected the algorithm's performance. Thus, in the proposed system the parameter, *step*, was used which defined the adjacent windows overlapping ratio. The speed of algorithm depended on the *step* value. In other words, if we increase the *step*, the computation burden of algorithm is increased exponentially and vice versa. In this study the optimum of *step* was selected 4 pixels for three databases empirically.

After generation of the sub-images, RT was applied to the each sub-image. High contrast between the ONH and image background can be resulted in generation of a peak in Radon space related to that sub-image which contains the ONH (if an ONH lays in the sub-image). In order to eliminate the diagonal effect, the input sub-images were firstly masked using a circle mask (like procedure shown in Fig. 7). After that the Radon peak was detected in the Radon space. Profiles in which peak occur were candidates that might contain the ONH. These profiles were further analyzed for validation of the ONH candidate.

The main concept of next part is similar to peak measuring in all projection angles in Radon transformation of a sub-image with a central point pattern. By this simple assumption, all sub-images which have a peak profile higher than a predefined threshold were taken into account.

To detect correct ONH, we used RT property for round objects. For a round object, RT has same profiles in all directions. Due to roundness of the ONH, profiles related to projections have minimum deviation with each other; therefore, we can detect that sub-image contains ONH with this procedure. To achieve this goal, mean square error between projections was calculated. In other words, we found the sub-image which minimized the mean square error between all its different projections.

After validation of the ONH by the described method, the center of sub-image which contained the ONH was considered as the center of a mask for the ONH. Since the accurate shape of the ONH was not important, the masked sub-image was just used in this study to identify of the ONH. (See Fig. 5)

MAs near the ONH were also masked, but masking a small area of retina, which may contain MAs, was not a very problematic issue due to occupation of these areas by the ONH. The complete procedure of ONH detection is explained in [30].

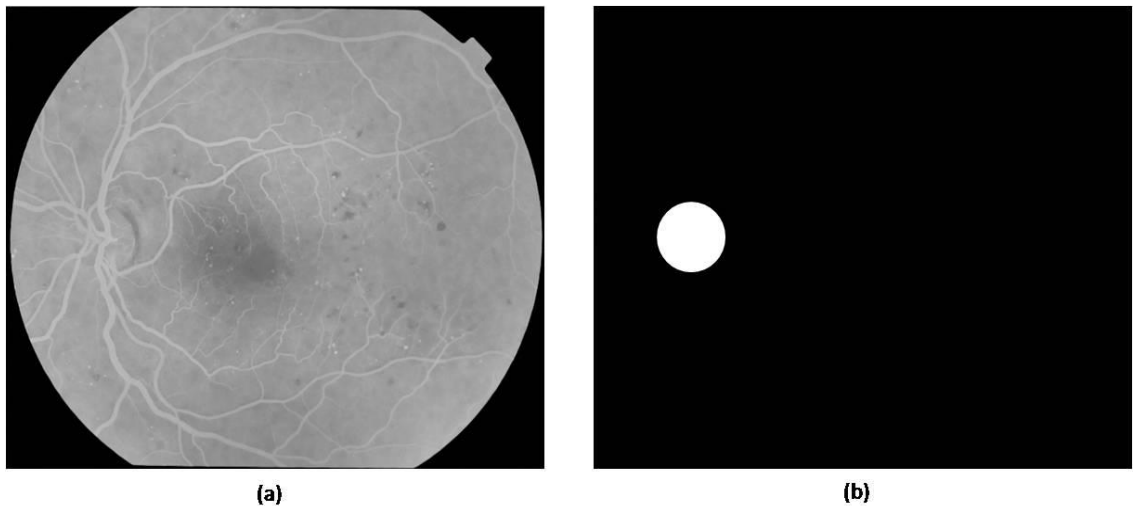

**Fig. 5.** (a) Original image (b) ONH detected and masked.

### 3.4. Preprocessing

Before searching abnormal lesions in an acquired fundus image, the image has to be preprocessed to ensure adequate level of success in detection of the abnormality. Preprocessing steps help to decrease background variation (in the first step) and eliminate MAs' like point noise (in the second step). The two steps of preprocessing were used in this project, including application of top-hat transformation and filtration of the image.

A top-hat transformation based on a disk ''structuring element'' whose best diameter ($d$=25 pixels) was empirically found to be the best compromise the complete segmentation between the features and background. The disk diameter depended on the input image resolution. After top-hat transformation, contrast stretching was used to increase contrast difference between features and background.

To eliminate point noise which could be falsely detected as MAs, an averaging filter was applied on the result of the top-hat transform (Averaging window size=75 pixels). Finally, we subtracted the result of averaging filter in the second step from the applying top-hat transformation in the first step. (Fig. 6)

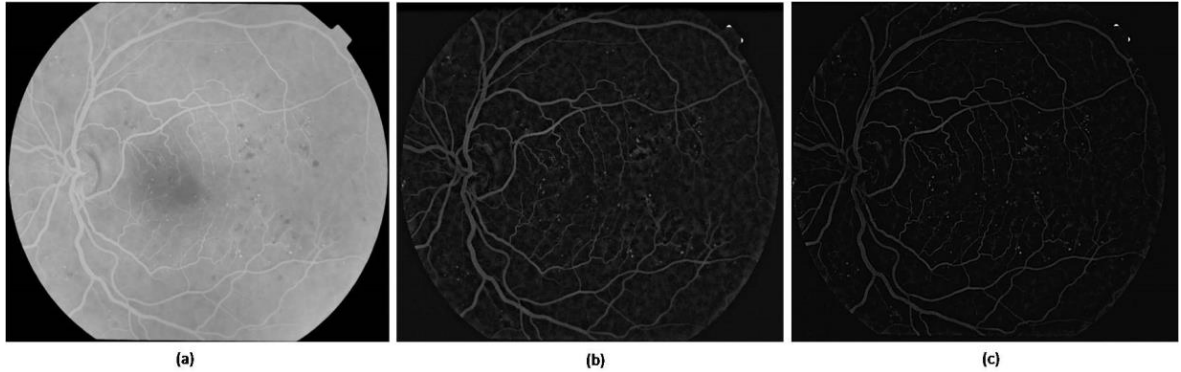

**Fig. 6.** (a) FA fundus image (b) top-hat result and contrast stretching (c) result of subtraction of top-hat and filtered top-hat image.

### 3.5.    *Main processing*

The results of preprocessing stage were utilized as input images for main processing. After preprocessing, vascular network and MAs are presented as bright features against darker background. To detect MAs, it is necessary to mask vascular tree, so remaining features would be mainly including MAs. Masking retinal vessel has positive effect on our results meaning that without detecting and masking vascular tree the result of our algorithm decreases intensely. Therefore, first of all, we detected retinal vessel network by using RT and multi-overlapping windows consequently, we detected MAs again by RT and fine level of thresholding.

#### 3.5.1.   *Detection of retinal vessels*

Blood vessels can be described as bright curvilinear objects against a darker background with ill-defined edges. The algorithm was composed of 4 steps:

1.  **Sub-images generation**

**2.** Radon transform

**3.** Candidate vessel certifying

**4.** Vessel refinement

In our proposed algorithm, FA fundus image was firstly partitioned into some overlapping windows. The size of window, $n$, was selected bigger than width of the biggest vessel in pixel ($n$=30 pixels for MUMS-DB, $n$=15 pixels for 2[nd]-DB, and $n$=17 pixels for ROC database). The windows overlapping ratio in the proposed system, *step*, has effect on the computation burden of algorithm like ONH detection. In this study the optimum of *step* was selected 5 pixels for both databases empirically. RT was then applied to each window. Because of the existence of more diagonal pixels than other directions, the amplitude of projection in diagonal directions ($\theta$=45°, $\theta$=135°, $s$=$n\sqrt{2}$) is higher than other directions, so the peak of RT is more likely to happen in diagonal directions. To eliminate the diagonal effect, a circular mask was multiplied with input sub-image. The masking process is shown in Fig. 7.

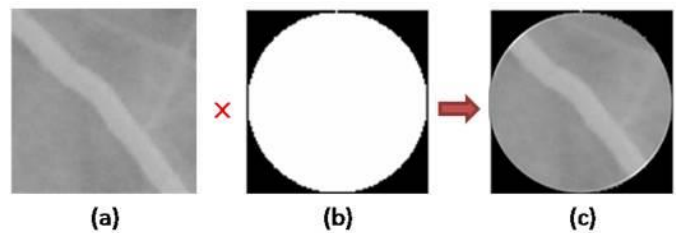

**Fig. 7.** Masking process: (a) Original sub-image (b) applied mask (c) masked sub-image.

The peak amplitude was compared with a predefined threshold. If the peak amplitude was bigger than threshold, the detected sub-vessel was confirmed and the algorithm should calculate sub-vessel's width. (Fig. 8)

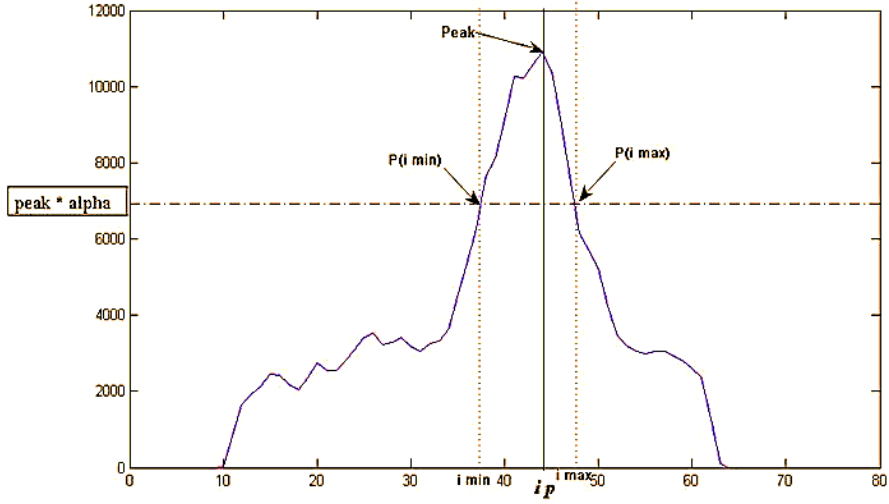

**Fig. 8.** Profile of Radon Transform in θ= 135° from Fig. 7.

In Fig. 8, $i_p$ is the peak's index associated with the sub-vessel's center line. The interval $[i_{min}, i_{max}]$ around $i_p$ is the sub-vessel's thickness.

$$i_{max} > i_p > i_{min} \qquad (3)$$

$$P(i_{max}) = \propto \times P(i_p) \qquad (4)$$

$$P(i_{min}) = \propto \times P(i_p) \qquad (5)$$

$$w = i_{max} - i_{min} + 1 \qquad (6)$$

In equations (5) and (6), $\alpha$ is a constant (0<$\alpha$<1) and $P$ denotes profile. The $\alpha$ was selected 0.75 for MUMS-DB and 0.6 for 2nd-DB and ROC database.

Sub-vessel mask was prepared by using $i_{max}$, $i_{min}$, projection angle ($\theta$), and window size ($n$) which presents coarse sub-vessel's coverage area in the sub-image.

For this purpose, on a black $n \times n$ block a white line was drawn which its angle equals to projection angle $\theta$ and its position was determined by $i_{min}$ and $i_{max}$ and width of drawn line determined by $w$. (Fig. 9d)

The output local vessel mask is not a good representation (Fig. 9d), and this phenomenon decreases the quality of finally detected vessel map and also disturbs the extracted statistical information. To overcome this problem and improve the quality of extracted sub-vessels in the sub-image, the local vessel mask should be compared with real sub-vessels. For this reason, the input sub-image is converted to a binary image which white pixels represent vessel area and black pixels represent background. (See Fig. 9(c)). After that two gray level means were computed. The first one was the mean of those pixels in preprocessed sub-image, located on white ribbon of the local vessel mask, and the second one was the mean of those pixels' gray level, which their corresponding pixels in the local vessel mask located on black background. To improve the quality of extracted sub-vessels in the sub-image, the local vessel mask was compared with real sub-vessels. Consequently, according to values of these two means, sub-image was thresholded to make a binary sub-image in which white pixels represented vessel area and black pixels was background. White and black pixels represented vessel area and background respectively. Fig. 9 shows the complete procedure of vessel refinement.

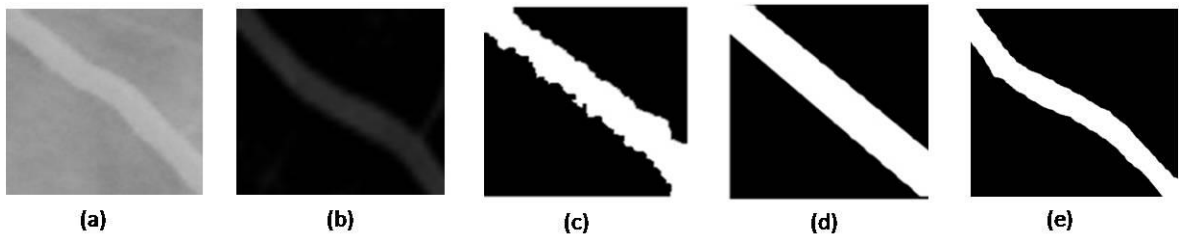

**Fig. 9.** (a) Original sub-image; (b) preprocessed sub-image; (c) binary sub-image; (d) local vessel mask; (e) fine local mask of sub-image.

Eventually, to achieve the vessel map of the input image, fine local masks were merged. In order to merge fine local masks, adjacent masks were put beside each other considering the overlapping ratio; on the overlapping region a logical *OR* was performed. (See Fig. 10) This method is completely explained in [31]. The results of vessel detection are shown in Fig. 11, Fig. 12 and Fig. 13.

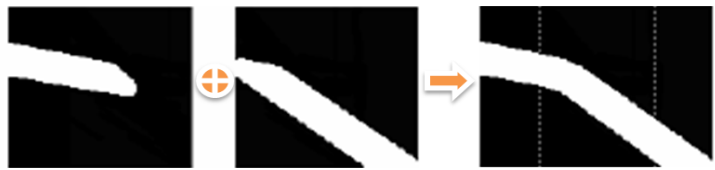

**Fig. 10.** Result of logical OR of sub-images

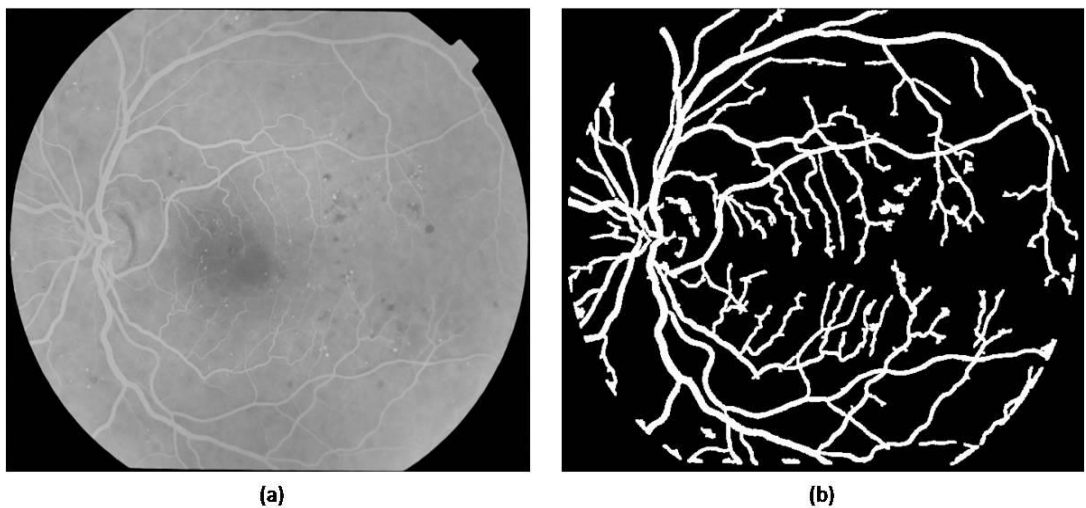

**Fig. 11.** Vessel detection result from MUMS-DB, (a) input image (b) vessel mask.

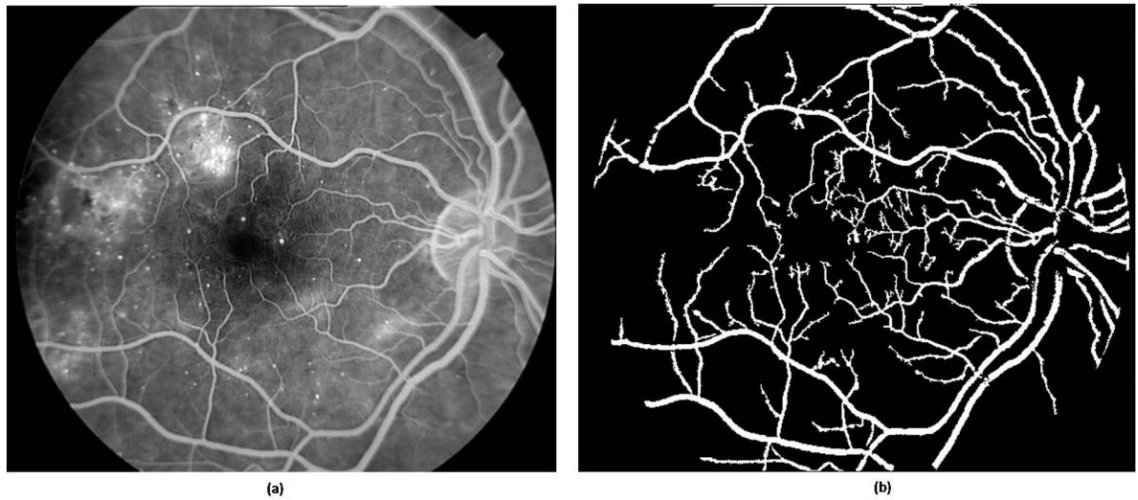

**Fig. 12.** Vessel detection result from second local database, (a) input image (b) vessel mask.

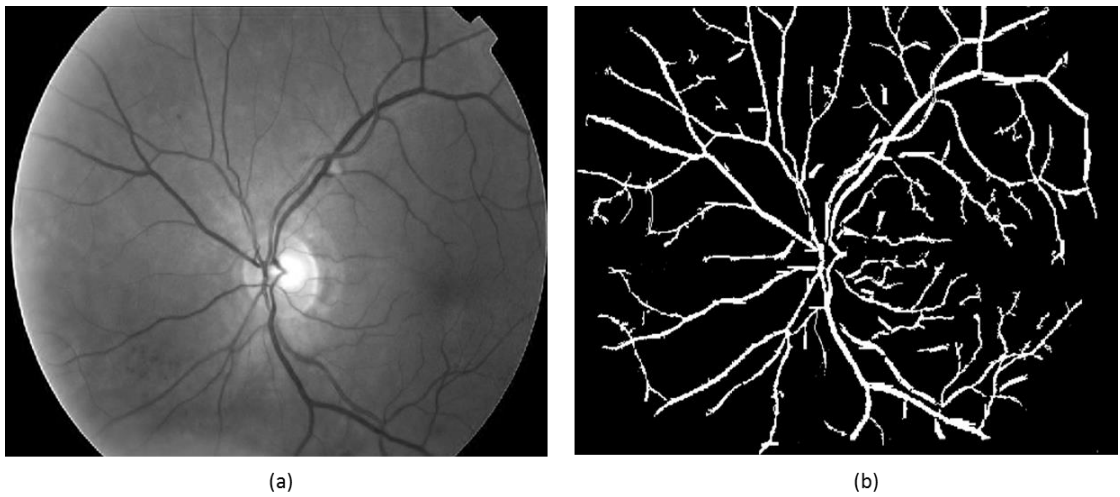

**Fig. 13.** Vessel detection result from ROC database, (a) input image (b) vessel mask.

### 3.5.2. Detection of Microaneurysms

To eliminate all interfering effects, we masked our images with two different masks; the first one was the ONH mask and the second one was the vascular tree mask.

In this section a RT Based Approach for Extraction of circular pattern not only was utilized for MA detection but also simplified the statistical analysis of the input retinal image. In order to extract MA, circular pattern with diameter lower than 125 µm should be extracted in local windows. The window size (*n*) has a direct effect on the extraction accuracy. A small/large *n* would lead to extract small/big semi-circular pattern accurately, while big/small lesions would not be extracted. Good lesion detection can be achieved by adjusting the *n* value. The size of *n* was selected equal to maximum diameter of the biggest MA in pixel. Here we found *n*=18 for MUMS-DB, *n*= 10 for 2nd-DB, and *n*=12 for ROC database empirically.

Another important parameter which affected the algorithm's accuracy was the windows' overlapping factor, in that non-overlapping windows extremely limit the quality of detected lesion. Thus, in the proposed system, a parameter, *step*, was used which defined the adjacent windows overlapping ratio. Same as the ONH and vessel detection the *step* depended on the speed of algorithm. If we increase the *step* the computation burden is increased exponentially and vice versa. Here the optimum measure of *step* we found was 5 pixels empirically. After applying RT, the peak of RT in Radon space would be associated with the circular pattern (if a MA lays in the sub-image). To eliminate the diagonal effect, masking process has to be done on each sub-image using a circle mask before applying RT. Masking is shown in Fig. 14.

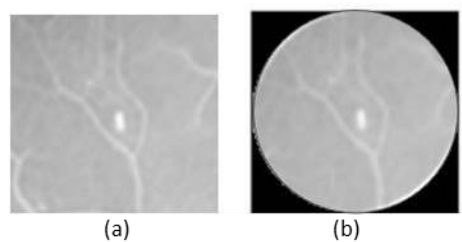

(a)  (b)

**Fig. 14.** (a) Sub-image, (b) masked sub-image.

The window, which includes the pre-defined peak was further analyzed for MA validation. high contrast between MA and the sub-image background can be resulted in generation of a peak for the MA in Radon space. The sub-image, in which peak occured, was a candidate that might contain a circular pattern. If MA located in the sub-image then the peaks would be found in every columns of Radon space, it means that in all projection angle we should see a Gaussian pattern in column, with sharp peak in the rows. Another important issue was semicircular pattern of MA, which was made Gaussian peaks approximately similar together in all projection angle. For a round object, RT has same profiles in all directions. Due to roundness of the MAs, profiles related to projections have minimum deviation with each other.

On the pther hand, Point noise, end point of vessels, and bifurcations are similar to MAs (false MAs) (Fig. 15). Therefore, to validate MAs, following MAs characteristics in the images were used:

- Intensity: They have luminous isolated peaks, i.e. they are much brighter than the background and they are disconnected from the network of blood vessels.

- Size: Their diameter is less than 125 μm.

- Shape: Their shape can be estimated by a two-dimensional Gaussian [23].

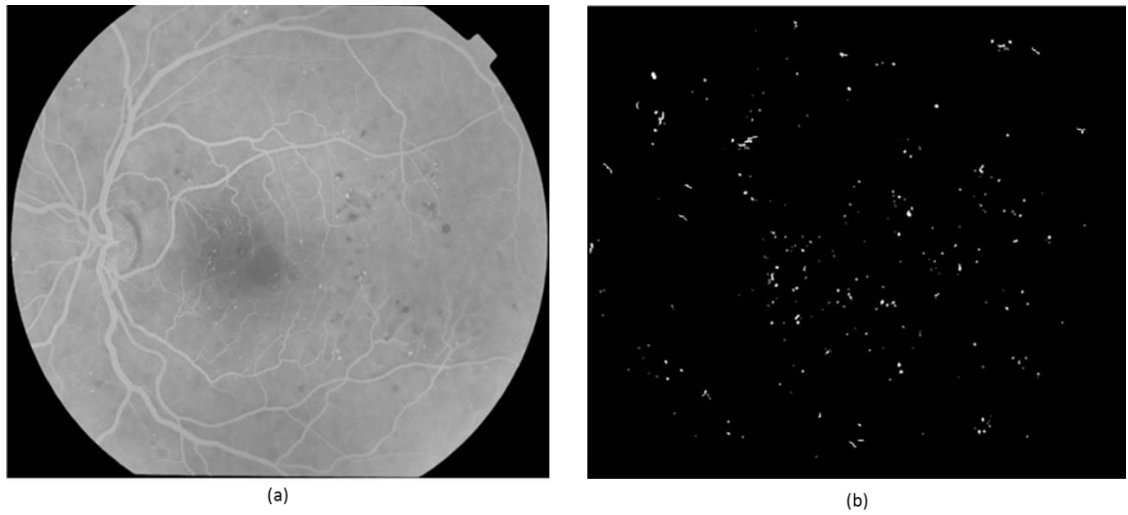

(a)    (b)

**Fig. 15.** (a) Input image from the MUMS-DB (b) result of MAs detection with false and true MAs

Thresholding is a way to evaluate the intensity. In other words, an easy solution to the MA validation problem is to compare the peak amplitude with predefined thresholds. Size and shape of candidate were checked using RT on each candidate. Circular shaped MA candidate with diameter lower than 125 μm extract from final output binary image. Semi-circular pattern like MA create this peak in all columns of the RT matrix. (Fig. 16, 17 and 18)

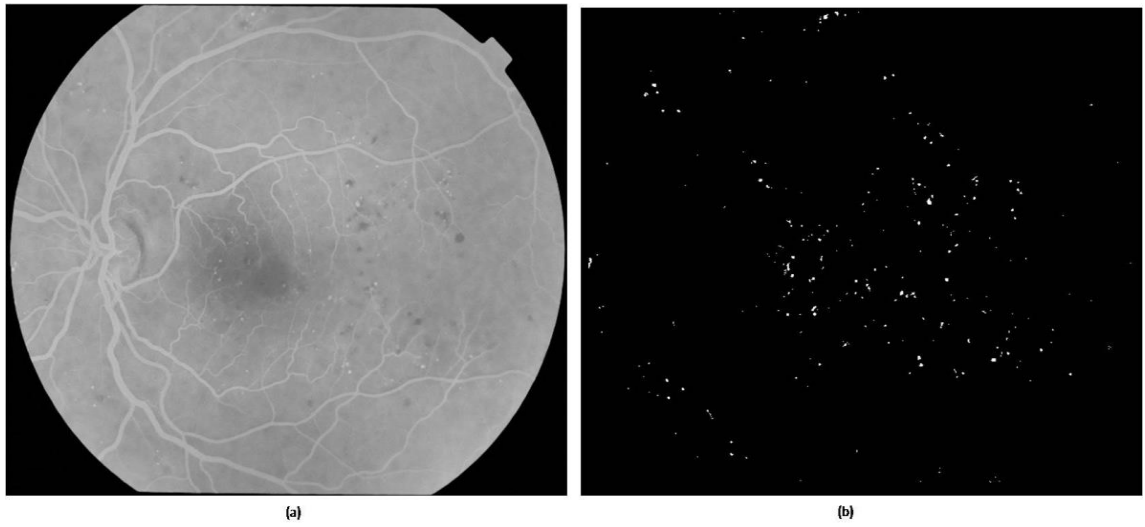

**Fig. 16.** (a) Input image from the MUMS-DB (b) final result of MAs detection.

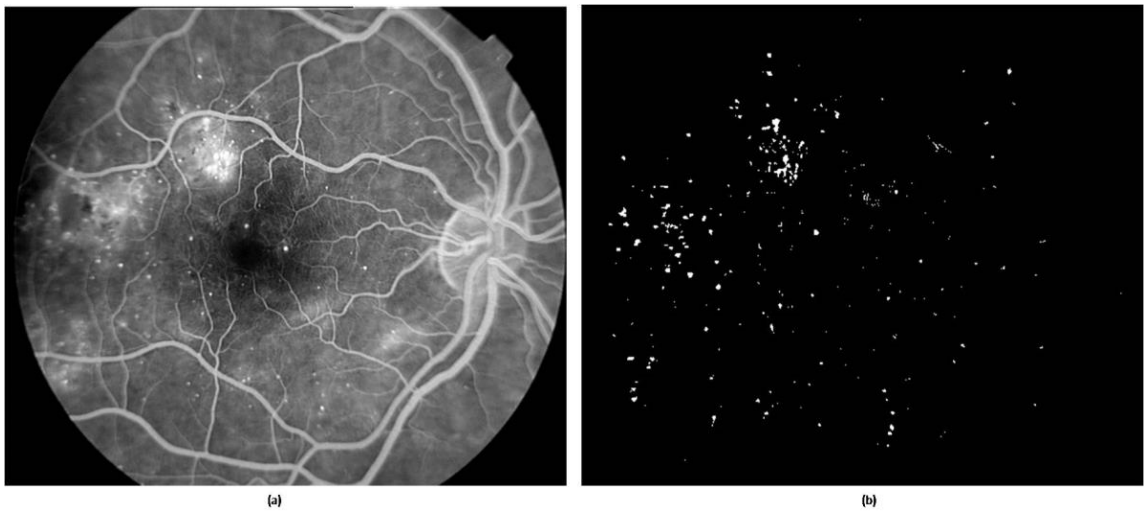

**Fig. 17.** (a) Input image from the 2ⁿᵈ-DB (b) final result of MAs detection.

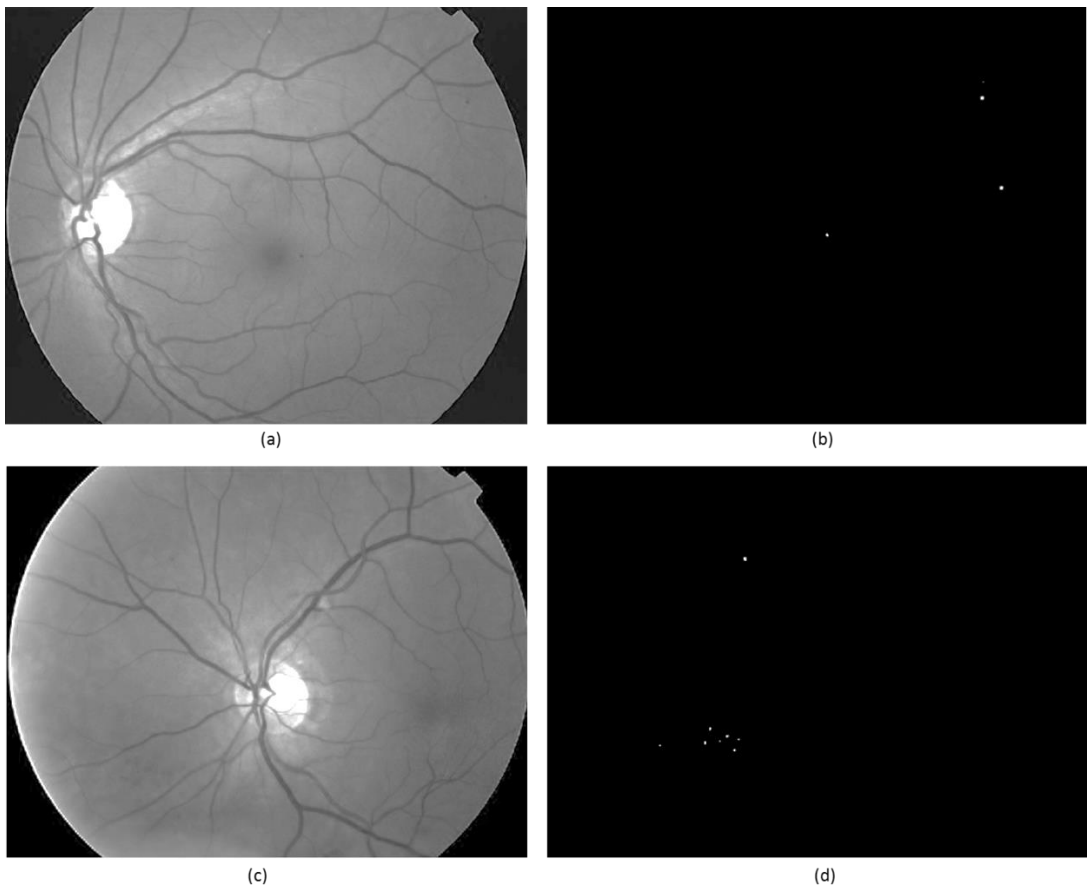

**Fig. 18.** (a), (c) Input images from the ROC database (b), (d) final result of MAs detection.

By using appropriate preprocessing and also acurate vasculare detection method, automatic system could mask some parts of vessels that were assumed MAs (false MAs).

## 4. Experimental Results

Patient or image and the pixel or lesion based analysis were used to calculate sensitivity and specificity of proposed method. In an automated program, the accuracy of patient based analysis is more important from the clinical point of view. Pixel based analysis would provide more information on accuracy of the proposed method.

On the other hand, using a computer with 2.0GB RAM and 2.33GHz Core 2 Due CPU, processing of one image took around 11 minutes for MUMS-DB [665.20 ± 80.368 seconds] including pre-processing [37.56 ± 1.308 seconds], Vessel detection [580.37 ± 33.814 seconds] and MA detection [41.66 ± 9.128 seconds].

### 4.1. *Sensitivity and Specificity*

The accuracy of the diagnosis was assessed using the sensitivity and specificity measures. Sensitivity means the percentage of abnormal funduses classified as abnormal by the procedure. Specificity means the percentage of normal funduses classified as normal by the procedure. The higher the sensitivity and specificity values, the better the procedure. Sensitivity and specificity values can be calculated as follows.

$$\begin{cases} Sensitivity = \dfrac{TP}{TP + FN} \\ Specificity = \dfrac{TN}{TN + FP} \end{cases} \qquad (7)$$

In which *TP*, *TN*, *FP*, and *FN* mean true positives, true negatives, false positives, and false negatives, respectively.

In image based analysis, whenever there are more than 5 MAs in an image, DR would be diagnosed. Positive/negative value reflects the result of automated diagnosis of DR regarding to expert ophthalmologist diagnosis in form of true/false.

In pixel or lesion based analysis, positive/negative values are related to automated diagnosis of pixels belong to MA or not in comparison to precise position, marked by ophthalmologists and True/false values are related to pixels which may belongs to MAs or not in ground truth file.

### 4.2. Training and test set

In this study, we used 45 images for a training set (for learning). This consisted of 35 images from the MUMS-DB and 10 images from the $2^{nd}$-DB. Moreover, we applied 2 images from ROC database in this database. The test set (test purposes) consisted of 145 images of which 85 images from the MUMS-DB, 40 images from the $2^{nd}$-DB, and 20 images from the ROC database. After fixing the parameters of our algorithm by using training set, automated method was tested in each image of our three databases.

In automated MA detection one of the main problems is to establish a ground truth file. In order to determine ground truth, we asked an expert ophthalmologist with more than 15 years in translating retinal images and a trained ophthalmologist to locate MAs in the images from train and test set independently. We fixed a protocol letting neither image manipulation nor zoom. For every doubtful candidate or those that they disagree with each other, we asked these two ophthalmologists to discuss their results and to find an agreement. The result of this discussion was considered as a ground truth to which our method could be compared.

In order to estimate the performance of the ophthalmologists (and thus the difficulty of finding a gold standard), we compared the two independent readings to the ground truth files. The results are shown in Table 1.

**Table 1**

Comparison between two readers and gold standard

|  | Reader A | Reader B |
|---|---|---|
| Sensitivity | 97.8% | 86.9% |
| FP per image | 0.41 | 2.58 |

### 4.3.    Detection of vascular tree

Small vessels can appear as succession of small isolated patterns, which could be detected as FP in detection of MA. Vascular detection, in the first step of main processing, removed the vascular tree map from the retina. The performance of the main step of this study, the detection of MAs, depends on the accuracy of vessel masking. Fig. 20 and Fig. 21 show the result of MA detection in our two FA databases with and without vessel removal.

Sensitivity did not change in image based analysis for detection of DR in the MUMS-DB and $2^{nd}$-DB, and specificity is decreased from 75% with vascular removal to 61% without it in the MUMS-DB and from 70% to 56% in the $2^{nd}$-DB. In lesion based analysis without vascular masking, sensitivity increases from 92% to 98.3% in the MUMS-DB, and increases from 95% to 100% in the $2^{nd}$-DB.

In first situation (with vascular removal) we have 5 MA FP per image in MUMS and $2^{nd}$ databases while without vascular removal it increases to 10 FP per image.

### 4.4.    Detection of Microaneurysms

### 4.4.1.  Image based analysis

Of the 85 images of the MUMS-DB test set, the algorithm gave 66 positive diagnoses (more than 5 MAs) for DR, consisting of 61 TP, and 5 FP in comparison to the ground truth image. Between healthy fundus images, 15 out of 20 images were diagnosed truly (TN), while there were 4 image falsely diagnosed as healthy (FN). So the algorithm has a sensitivity of 94% and specificity of 75 % at the level of image based analysis.

The algorithm diagnosed 33 images with DR, consisting of 30 TP and 3 FP of all 40 test images from the $2^{nd}$-DB (30 images with DR and 10 normal). In the case of healthy images, 7 out of 10 images were diagnosed truly. Therefore, in detection of DR, the algorithm reached to sensitivity and specificity of 100% and 70% respectively.

**Table 2**

Results of image base MA detection in the MUMS-DB

| Test results | TP | FN | TN | FP |
|---|---|---|---|---|
| Number of images | 61 | 4 | 15 | 5 |
| Sensitivity | $\frac{61}{61+4} = 94\%$ | | | |
| Specificity | $\frac{15}{15+5} = 75\%$ | | | |

**Table 3**

Results of image base MA detection in the $2^{nd}$-DB

| Test results | TP | FN | TN | FP |
|---|---|---|---|---|
| Number of images | 30 | 0 | 7 | 3 |
| Sensitivity | $\frac{30}{30+0} = 100\%$ | | | |
| Specificity | $\frac{7}{7+3} = 70\%$ | | | |

The Receiver Operating Characteristic (ROC) curve (Fig. 19) of the automated MA detection illustrates the range of sensitivity and specificity in detection of DR in the MUMS and $2^{nd}$ databases. The Area Under the Curve (AUC) of the ROC curves was 94.8% for the $2^{nd}$-DB and 90.5% for the MUMS-DB.

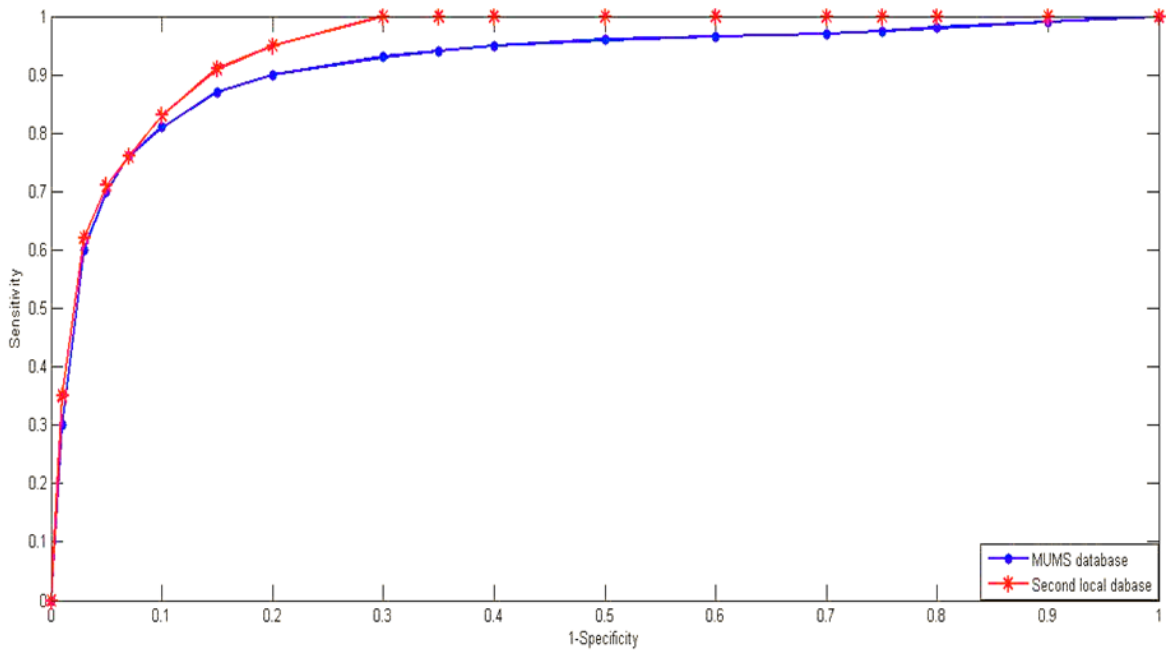

**Fig. 19.** ROC of automated diabetic retinopathy MA detection algorithm for two databases.

### 4.4.2. Lesion based analysis

In this section both the algorithm efficiency in detection of the MAs and the spatial precision of the results were evaluated. The evaluation in this section was lesion based rather than pixel based.

According to ground truth, the number of MAs in all 85 images from the MUMS-DB was 6710. The average of MAs for each image was 79. Our automated method found 6848 MAs, which 6160 MAs of the ground truth were diagnosed truly (TP), and 550 MAs were missed (FN). 688 out of 6848 detected MAs were located in healthy tissue of retinal images according to ground truth files (FP). Therefore, sensitivity of method was 92% while calculation of the specificity in lesion or pixel based fashion needs the true negative value. We should look on healthy retinal pixels marked healthy, but masking process in MA detection method make this evaluation meaningless. Another statistical value of help in evaluating a method is the average FP per image value; this value was equal to 5 MAs.

In the 2$^{nd}$-DB, the number of MAs in all of 40 images was 907 (the average of 23 MAs for each image). Automated method, found 1059 MAs from which 859 MAs of the ground truth were diagnosed truly (TP), and 48 MAs were missed (FN). 198 out of 1059 detected MAs were located in healthy tissue of retinal images according to ground truth files (FP). Sensitivity of method was 95%. In other side, FP per image value was equal to 5 MAs.

In ROC database, between 135 MAs, our automated method found 122 MAs truly (TP), and 13 MAs were missed (FN). Moreover, the FP was 41 and the sensitivity of method was 91%. In other side, FP per image value was equal to 5 MAs.

**Table 4**

Sensitivity of lesion base MA detection from the MUMS-DB

| Test results | TP | FN | TN | FP |
|---|---|---|---|---|
| Number of images | 6160 | 550 | --- | 688 |
| Sensitivity | $\frac{6160}{6160+550} = 92\%$ | | | |

**Table 5**

Sensitivity of lesion base MA detection from the second database

| Test results | TP | FN | TN | FP |
|---|---|---|---|---|
| Number of images | 859 | 48 | --- | 198 |
| Sensitivity | $\frac{859}{859+48} = 95\%$ | | | |

**Table 6**

Sensitivity of lesion base MA detection from the ROC database

| Test results | TP | FN | TN | FP |
|---|---|---|---|---|
| Number of images | 125 | 10 | --- | 41 |
| Sensitivity | $\frac{122}{122+13} = 91\%$ | | | |

For comparison our results in this part, the most common way of reporting algorithm performance is Free-response Receiver Operating Characteristic (FROC) analysis. In FROC analysis the sensitivity of MA detection system was plotted against the average number of FP detections per image.

The performances of algorithm with and without vessel masking in MAs detection for MUMS and 2nd databases are illustrated in Fig. 20 and Fig. 21. Moreover, in Fig. 22 the comparison between the three databases in MA detection is shown.

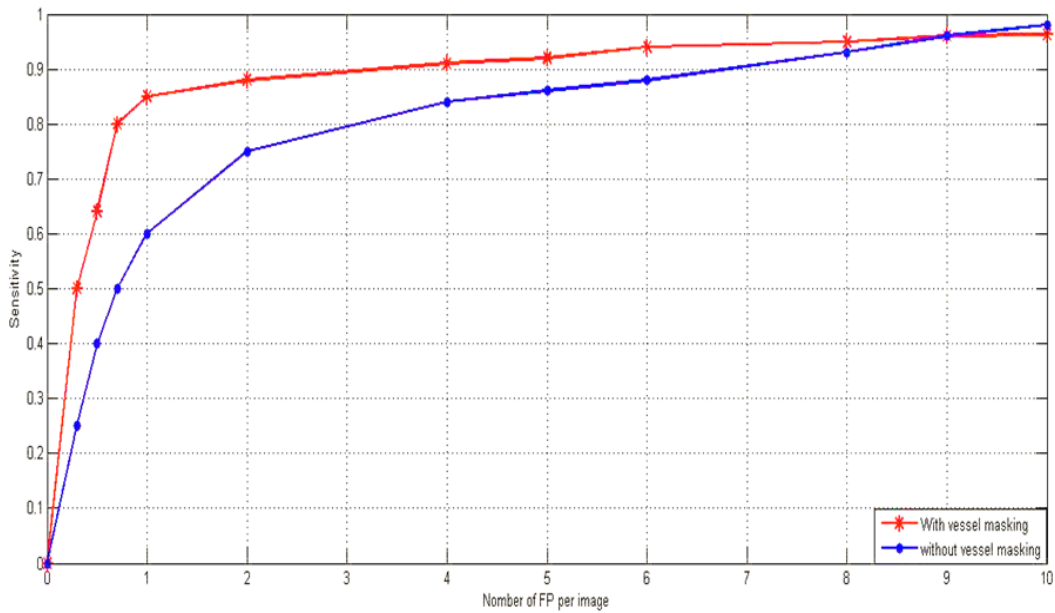

**Fig. 20.** FROC for our algorithm with and without vessel masking in the MUMS-DB.

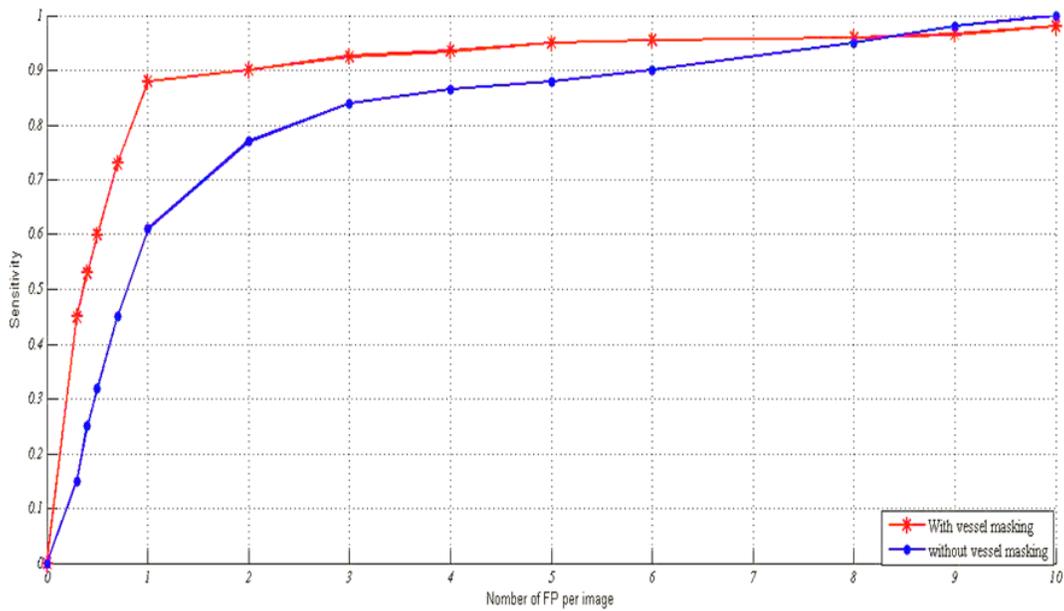

**Fig. 21.** FROC for our algorithm with and without vessel masking in the 2nd-DB.

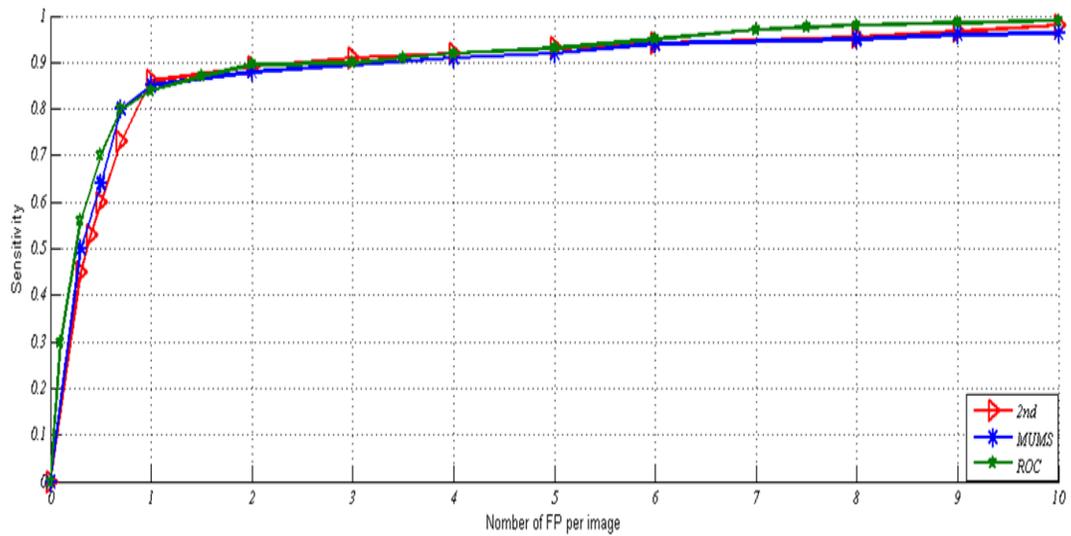

**Fig. 22.** FROC for our algorithm with vessel masking between the three Databases.

### 4.5. Discussion on MAs results

Our results were as good as some previous reports using FA. Cree *et al.* [24] reported 82%

sensitivity and 84% specificity and 5.7 false-positives per image, although there was less MAs per image in his study compare to our data set (MAs per image =79). Spencer *et al.* [21] reached to 82% sensitivity and 86% specificity because of missing 13 MAs in the candidate-microaneurysm images.

In comparison with other MA detection methods, our method has some similarity to Hipwell *et al.* because of processing steps using top-hat transformation and specially vascular removal before MA detection, its produced a sensitivity of 81%, with 93% specificity [32] against the reference standard of an experienced clinical research fellow according to the EURODIAB HMA protocol [33].

Walter *et al.* [23] tested the algorithm on 23 images; their algorithm detected nearly 80% of the MAs pointed by the ophthalmologist. However, there were also false positives, mainly on the papilla and on points of laser treatments. But they didn't declare what the specificity was. On the other hand, same as Walter *et al.* [23] and Mendonça *et al.* [22] we used some characteristics for validation of MAs such as local intensity and shape. Moreover, To segment the MA from other retinal features, Hafez and Azeem used the size, shape, and energy characteristics analysis [27].

In addition, in Spencer *et al.* [21], Mendonça *et al.* [22], and Frame *et al.* [34] variations of Lay's method were discussed. On the other hand, Niemeijer *et al.* [13] worked on MA detection in color fundus images and in order to avoid missing MAs, they combined their method with Spencer *et al.* [21] and Flame *et al.* [34].

## 5. Discussion

Detection of MAs is only a single step in early detection of DR. Although MAs can be seen in color fundus photographs, they are best visualized on FA due to enhanced contrast. It is not only more difficult to diagnose MAs in the red background of the color fundus images, but also differentiation between them and small hemorrhages is often difficult or even impossible. Therefore, FA is considered as gold standard procedure in diagnosis of MAs.

Many frames are taken during the process of FA. From computer processing techniques point of view, the best quality frame is chosen from those frames that are captured immediately after complete filling of the vasculature.

Vascular detection is the first step of main processing which removes the vascular tree map from the image. The performance of the main step of this study, the detection of MAs, depends largely on the accuracy of vessel detection. In vessel segmentation, proposed method was tested for the first time on DRIVE database by Pourreza *et al*. [35] and their results have been published. They also checked out on vascular tree of conjunctiva and FA retinal images by same research group in the MUMS [31, 36].

Obviously, the specificity of MA detection in proposed method will decrease if it works on images with poor vascular detection.

As we said before, in looking toward an integrated program for automated detection of DR it is important to have adequate sensitivity. If we look on the whole project, the patient base analysis of MA detection, especially increasing sensitivity of this section, is the most important goal of this study. This indicates that while finding individual MAs is critical in detection of those patients with DR, it may not be necessary to detect every individual MA in all images for DR detection. The sensitivity of 94% for the MUMS-DB and 100% for the $2^{nd}$-DB makes our CAD as good as ACD in diagnosis of DR.

On the other hand, although DR detection systems are evaluated on the overall presence of DR in the patient, not the number of MAs in a single image, we also focused at the lesion level, because we do not present a whole system of automatic diagnosis, but only a part of it; therefore, it is most informative to see how good the detection of these lesions really is. In this section in addition to assess the efficiency of our method to detect the MAs, the spatial precision of the results were evaluated. The MA detection method reached to sensitivity of 92% for MUMS-DB, 91% for ROC database, and 95% for $2^{nd}$-DB, in lesion based analysis that showed the ability of our algorithm even in follow up and treatment planning of DR. Furthermore, we chose the use of FROC analysis for statistical analysis. The FROC analysis also lacks an AUC summary variable in that ROC analysis does have. Our purpose to choose FROC over other evaluation measures was that it is generally regarded as the evaluation measure most closely resembling clinical practice in diagnostic tasks that involve lesion localization [7] .

Meanwhile, our FN seems to be due to: (1) in some cases some MAs were located adjacent to vessels and were detected as vessels (2) some MAs that were near each other or some were too big were also detected as vessels.

Our designed algorithm was robust against lower image resolution. We intentionally used non selected images to check our method in real work atmosphere, including poor quality images and also post-treated retina with coagulated points some of FP MA detection in this study were due to these coagulated points on retinal images.  In our study, there were two other sources of FP diagnosis: vascular cutoff and telangectasia.

Image capturing took around 2-3 minutes depending on the photographer's experience and also patient cooperation and expert ophthalmologist could read every image in less than 6 minutes. Total time consuming in our method (11 minutes for MUMS-DB) was a little more than the manual time but the automated system could work more, without tiredness and decreasing accuracy due to it.

Beside using RT as in feature extraction method in retinal FA, another important issue in this study was the use of multi-overlapping sub-images to prevent error caused by global techniques. In global techniques, because of more background variation in contrast of small sub image analysis, error occurs more.

It is important the computer-based system used in an automated program have high sensitivity and no images with DR are missed.

One advantage of the automated systems is that they are deterministic. In other words, they always classify funduses in a similar way. There will always exist differences in the backgrounds and education of human experts which causes dispersion in their diagnosis. Moreover, a single human expert may make different diagnoses in different screening times due to human factors, such as tiredness or sickness. The performance of the human expert indicates the difficulty of the task. This also illustrates experts may have been less sensitive than some of the other experts in detecting lesions. It is better to use a computer-based system for classifying only clearly normal funduses as normal, whereas abnormal and obscure funduses are delivered to a human expert for further classification. However, the automated system reduces the workload of the human expert, since in reality most of the funduses are normal, and only a few funduses have retinopathy.

Therefore, if the first phase of the diagnosis is performed automatically by a computer, it is important to have as high sensitivity as possible so that no retinopathic funduses are missed. Also in the case of the combined computer and human maximum sensitivity and specificity values should be achieved.

## 6. Conclusion

Although the final purpose of this study was early detection of DR, this issue was only about detection of MAs. The goal of this work was to develop an algorithm for detection of different abnormal vascular lesions related to DR and development of a complementary automated detection of DR in Iran for first time. Altogether, the aim in retinal angiography is to investigate and extract as much information as possible. So to achieve this purpose, using a system with the highest image quality is critical.

Altogether, for all MAs the human observer is well ahead of the automated methods. The results proved that it is possible to use algorithms for assisting an ophthalmologist to segment fundus images into normal parts and lesions, and thus support the ophthalmologist in his or her decision making.

To utilize this program in the follow up of patients, we should add an image registration algorithm so that the ophthalmologist could study the effect of his/her treatment and also the progression of the disease, not only by crisp counting, but also by spatial orientation which is included in presented method.

**Acknowledgement**

The authors acknowledge support for a Master degree grant No. 87394 from Mashhad University of Medical Sciences (MUMS). Besides, Meysam Tavakoli would like to thank Dr. Faraz Kalantari, Dr. Alireza Fadavi Boostani, and Dr. Azin Nazar for their valuable suggestions in final preparation of this manuscript.